\documentclass[aps,prl,twocolumn,superscriptaddress,showpacs]{revtex4-1}
\usepackage[colorlinks,linkcolor=blue,citecolor=blue,urlcolor=blue,breaklinks=true]{hyperref}
\usepackage{amsmath,amssymb,graphics,epsfig,epstopdf,color,verbatim,braket,tabularx,breakurl}
\usepackage{multirow}
\usepackage{diagbox}
\usepackage{subfig}
\usepackage[normalem]{ulem}
\usepackage{float}
\usepackage{ragged2e}

\def\add#1{{\textcolor{black}{#1}}}
\def\fp#1#2{\frac{\partial#1}{\partial#2}}
\def\tred#1{{\textcolor{black}{#1}}}
\begin{document}

\title{The Green's function Monte Carlo combined with projected entangled pair state approach to the frustrated $J_1$-$J_2$ Heisenberg model}

\author{He-Yu Lin}
\address{Department of Physics, Renmin University of China, Beijing 100872, China}
\address{\tred{Key Laboratory of Quantum State Construction and Manipulation (Ministry of Education), Renmin University of China, Beijing, 100872, China}}
\author{Yibin Guo}
\address{CQTA, Deutsches Elektronen-Synchrotron DESY, Platanenallee 6, 15738 Zeuthen, Germany}
\author{Rong-Qiang He}
\address{Department of Physics, Renmin University of China, Beijing 100872, China}
\address{\tred{Key Laboratory of Quantum State Construction and Manipulation (Ministry of Education), Renmin University of China, Beijing, 100872, China}}
\author{Z. Y. Xie}
\email{qingtaoxie@ruc.edu.cn}
\address{Department of Physics, Renmin University of China, Beijing 100872, China}
\address{\tred{Key Laboratory of Quantum State Construction and Manipulation (Ministry of Education), Renmin University of China, Beijing, 100872, China}}
\author{Zhong-Yi Lu}
\email{zlu@ruc.edu.cn}
\address{Department of Physics, Renmin University of China, Beijing 100872, China}
\address{\tred{Key Laboratory of Quantum State Construction and Manipulation (Ministry of Education), Renmin University of China, Beijing, 100872, China}}
\address{\tred{Hefei National Laboratory, Hefei 230088, China}}
\date{\today}

\begin{abstract}
The tensor network algorithm, a family of prevalent numerical methods for quantum many-body problems, aptly captures the entanglement properties intrinsic to quantum systems, enabling precise representation of quantum states. However, its computational cost is notably high, particularly in calculating physical observables like correlation functions. To surmount the computational challenge and enhance efficiency, we propose integrating the Green's function Monte Carlo (GFMC) method with the projected entangled pair state (PEPS) ansatz. This approach combines the high-efficiency characteristics of Monte Carlo with the sign-free nature of tensor network states and proves effective in addressing the computational bottleneck. To showcase its prowess, we apply this hybrid approach to investigate the antiferromagnetic $J_1$-$J_2$ Heisenberg model on the square lattice, a model notorious for its sign problem in quantum Monte Carlo simulations.
Our results reveal a substantial improvement in the accuracy of ground-state energy when utilizing a preliminary PEPS as the guiding wave function for GFMC. By calculating the structure factor and spin-spin correlation functions, we further characterize the phase diagram, identifying a possible columnar valence-bond state phase within the intermediate parameter range of $0.52 \textless J_2/J_1 \textless 0.58$. This comprehensive study underscores the efficacy of our combined approach, demonstrating its ability to accurately simulate frustrated quantum spin systems while ensuring computational efficiency.	

\end{abstract}

\maketitle
{\em Introduction.} Investigating the intricate and novel physics in quantum many-body systems is a thriving research field in contemporary condensed matter physics. Given the limitations of computational resources, achieving a precise representation of the quantum state is crucial for addressing these challenges. Among the various approaches, tensor network states (TNS) are drawing increasing attention as they offer faithful representations of lowly-entangled quantum states, are free of sign problems, and have been successfully applied to a range of strongly correlated systems over the past decades \cite{verstraete2008matrix,Mont2018,orus2019tensor,cirac2021matrix,banuls2023tensor,xiang2023density}.

However, the associated computational cost of TNS escalates extremely fast with the maximum retained state number ($D$), commonly referred to as the bond dimension. This issue becomes particularly pronounced, especially in computing the physical observables, such as energy density and correlation functions. Furthermore, since most tensor network algorithms possess iterative structures \cite{orus2014practical,Roman2014EPJB,cirac2021matrix,xiang2023density,banuls2023tensor} that cannot be parallelized straightforwardly, the potential of existing acceleration architectures, including CPU multi-core parallelism and GPUs, cannot be explored easily.

On the other hand, Green's function Monte Carlo (GFMC) \cite{trivediGFMC}, belonging to the quantum Monte Carlo family, stands out as another frequently used numerical method in the study of strongly correlated many-body systems \cite{Anderson1975JCP,sorella2017book}. Starting from a prescribed trial wave function $|\Phi_0\rangle$, GFMC employs the fixed-node approximation \cite{Haaf1995PRB} to address the sign problem and leverages the imaginary-time evolution to refine its estimation of the true ground state \cite{Sorella2000PRB,Reynolds1982JCP,sorella2017book}. The physical observables, including the correlation functions, can be obtained efficiently by importance sampling, and the accuracy can be guaranteed as long as $|\Phi_0\rangle$ is provided with a certain degree of accuracy \add{ of the nodal information in particular.}

To take advantage of the high efficiency of Monte Carlo sampling and the sign-free feature of TNS simultaneously, in this study, we propose a hybrid approach \cite{MSL-PRB2000, Clark-arXiv2014, Sebastian-PRB2014, MPQin-PRB2020,CPB2022} that integrates the Green's function Monte Carlo (GFMC) method with the projected entangled pair state (PEPS) ansatz \cite{verstraete2004renormalization}, a specific type of tensor network state \add{which is expected to be able to capture the nodal information of the wave function}, to address the general quantum many-body problems. To illustrate the efficacy of this hybrid approach, we focus on the challenging frustrated $J_1$-$J_2$ Heisenberg model on a square lattice. This model is known for the possible existence of quantum spin liquid in the intermediate $J_2/J_1$ regime \cite{zhang2003valence,mezzacapo2012ground,jiang2012spin,Hu2013PRB,morita2015quantum,wang2018critical,ferrari2020gapless,liu2022gapless,Wang2013PRL,Liu2018PRB} but is difficult for quantum Monte Carlo simulations due to the severe sign problem \cite{SciAdv2020,pan2022sign}.

As expected, our computational results demonstrate a drastic improvement in the accuracy of ground-state energies, even when employing a preliminary PEPS as the guiding wave function for GFMC \cite{MPQin-PRB2020,CPB2022}. Furthermore, in comparison with the pure PEPS method, which needs to contract a tensor network with bond dimension $D^2$ in the expectation value calculations \add{(with leading cost $D^{12}$ generally)} \cite{orus2014practical, NTN2017, iTEBD2023}, the hybrid approach can fully use the highly efficient Markov chain and importance sampling techniques to expedite the calculations, and the resulting tensor network is of dimension $D$ only for each physical configurations (\add{with leading cost $D^6$} and thus can be contracted much more efficiently) \cite{LWang2011, WYLiu2017, HHZhao2017}.

Focusing on the intermediate phases, with periodic boundary conditions, to assess the validity and efficiency, we carefully benchmark the results obtained from the hybrid approach with those obtained from the exact diagonalization, density matrix renormalization group (DMRG), and some other approaches \cite{Schulz1994Magnetic,Hu2013PRB,Gong2014PRL,Choo2019PRB,Nomura2021PRX}. Besides the energy density, we also calculated the bond correlations and the static spin and dimer structure factors. Finally, we identify a possible columnar valence-bond state (VBS) phase \cite{AKLT1988} in the intermediate parameter regime about $0.52 \textless J_2/J_1 \textless 0.58$, \add{and we do not observe a possible nearby quantum spin liquid phase \cite{wang2018critical,Nomura2021PRX} in our calculations}.

{\em Model.} The frustrated $J_1$-$J_2$ Heisenberg model is defined by the following Hamiltonian,
\begin{eqnarray}
H = {J_1}\sum\limits_{\left\langle{i,j}\right\rangle}{{\mathbf S_i}\cdot{\mathbf S_j}} + {J_2}\sum\limits_{\left\langle\left\langle{i,j}\right\rangle\right\rangle}{{\mathbf S_i}\cdot{\mathbf S_j}},
\end{eqnarray}
where $\mathbf S_i$ is the spin-$1/2$ angular momentum operator defined at the $i$-th site on a square lattice, $\left\langle{...}\right\rangle$ and $\left\langle\left\langle{...}\right\rangle\right\rangle$ indicate the nearest and next-nearest neighboring summation, respectively. We focus on the case where both $J_1$ and $J_2$ are antiferromagnetic and consider system size $N = L \times L$ with periodic boundary conditions.

When ${J_1}$ dominates, the system is in a Neel phase with antiferromagnetic (AFM) long-range order \cite{Buonaura1998PRB, Sandvik1997PRB}, while
when ${J_2}$ dominates, the ground state manifests a well-established collinear AFM phase \cite{neel1936proprietes,neel1932proprietes}.
Nevertheless, the intermediate regime with $J_2/J_1$ around 0.5 remains a subject of considerable debate and scrutiny \cite{singh1999dimer,takano2003nonlinear,zhang2003valence,mambrini2006plaquette,darradi2008ground,murg2009exploring,mezzacapo2012ground,jiang2012spin,Hu2013PRB,morita2015quantum,wang2018critical,haghshenas2018u,ferrari2020gapless,hasik2021investigation,liu2022gapless,qian2024absence,Wang2013PRL,Liu2018PRB,zhitomirsky1996valence,PhysRevLett.84.3173,PhysRevB.79.024409,PhysRevB.85.094407,PhysRevB.89.104415,PhysRevB.44.12050,Nomura2021PRX,Gong2014PRL,Schulz1994Magnetic,Choo2019PRB,Zhang-PRB2024}. The strong quantum frustration and fluctuations pose a great challenge for numerical simulations. Despite numerous investigations into the characteristics of this regime, including the plaquette VBS \cite{zhitomirsky1996valence,PhysRevLett.84.3173,PhysRevB.79.024409,PhysRevB.85.094407,PhysRevB.89.104415,takano2003nonlinear}, the columnar VBS \cite{singh1999dimer,PhysRevB.44.12050,haghshenas2018u}, a gapless quantum spin liquid \cite{zhang2003valence,Liu2018PRB,Wang2013PRL,Hu2013PRB}, and other proposals \cite{mezzacapo2012ground,jiang2012spin,Gong2014PRL,morita2015quantum,wang2018critical,ferrari2020gapless,Nomura2021PRX,liu2022gapless,Zhang-PRB2024}, the precise nature of this quantum phase remains a topic of controversy. In this study, we always set ${J_1}=1$ for simplicity.

\begin{figure}[tb]
	\includegraphics[scale=0.8]{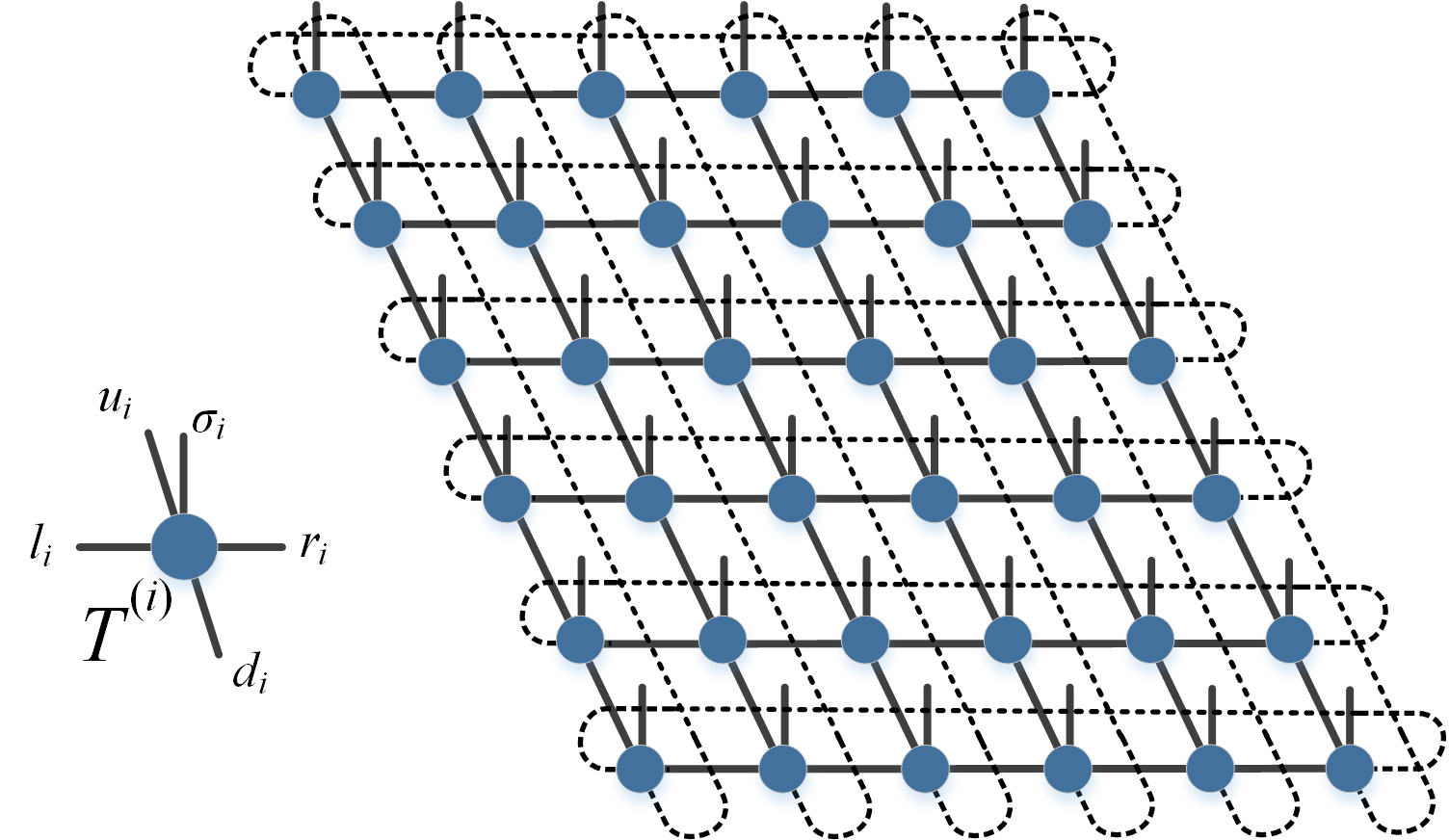}
	\caption{ \raggedright \tred{Sketch of the PEPS ansatz on a $6\times6$ square lattice with periodic boundary conditions (denoted as dashed). In our study, the local tensors $T^{(i)}$ of the PEPS wave function are optimized for $|\Psi_{2D}\rangle$ in the thermodynamic limit, as explained in detail in App.~II.}}
	\label{fig:PEPS}
\end{figure}

\emph{PEPS guiding wave function.}
PEPS \cite{verstraete2004renormalization} is a typical tensor network state extensively used to study two-dimensional quantum many-body systems. The PEPS ansatz we used in this study is sketched in Fig.~\ref{fig:PEPS} for $L = 6$ and can be formulated in the following
\begin{equation}
	\tred{|\Psi\rangle = \sum_{\{\sigma\}}\left[\mathrm{Tr}\prod_{i}T_{l_ir_iu_id_i\sigma_i}^{(i)}\right]\big|\sigma_1\sigma_2...\sigma_i...\big\rangle,
	\label{Eq:PEPS}}
\end{equation}
\tred{where $T^{(i)}$ is the local tensor defined at the $i$-th site with $(l_i,r_i,u_i,d_i)$ its link indices and $\sigma_i$ its local physical configuration, as show in Fig.~\ref{fig:PEPS}. For any given spin configuration $\{\sigma\}$}, the superposition coefficient is given by the trace (Tr) in Eq.~(\ref{Eq:PEPS}), which denotes the contraction of a two-dimensional tensor network, namely summation over all the link indices of the local tensors. The bond dimension $D$, the highest value that the link indices can take, is an important parameter in tensor network states. By increasing $D$, the number of parameters and the representation capability can be enhanced, but the computational cost escalates fast too \cite{orus2014practical, NTN2017}. Therefore, one should make a balance between performance and cost. \tred{In our calculations, the bond dimension we focused on is $D = 4$, but near the phase boundary, we have pushed to larger $D$ (no larger than 7) to check consistency. More background on tensor network states is provided in App.~I.}

This study considers the PEPS wave function as the trial ground state and guiding wave function for the GFMC calculations below. To generate such a trial state, we first perform energy minimization through \tred{variational approach with the help of automatic differentiation technique \cite{Biggs2000, Baydin2015}} for the two-dimensional Hamiltonian to get the ground state $|\Psi_{2D}\rangle$ in the thermodynamic limit \cite{dTRG-PRB2020, Schmoll2023,hasik2021investigation}, and then approximate the guiding wave function $|\Psi_0\rangle$ of the same Hamiltonian but on a $L\times L$ torus by placing the local tensors of $|\Psi_{2D}\rangle$ there, as shown in Fig.~\ref{fig:PEPS}. \tred{More details of the preparation of $|\Psi_0\rangle$ are explained in App.~II.}

\emph{GFMC method.} The basic idea of GFMC is simple \cite{Anderson1975JCP,sorella2017book}. Starting from an arbitrary state $|\Phi_0\rangle$, it performs the imaginary-time evolution to get the desired ground state $|\Psi_g\rangle$, that is
\begin{equation}
	\lim\limits_{\beta\rightarrow\infty}e^{-\beta \hat{H}}|\Phi_0\rangle \rightarrow |\Psi_g\rangle,
    \label{Eq:evolution}
\end{equation}
where the initial trial state satisfies $\langle\Phi_0|\Psi_g\rangle\neq 0$. In practice, the evolution process is divided into many small slices by setting $\beta = M\tau$ with $\tau$ a small number, and then the so-called Green's function is defined as the matrix element of $e^{-\tau \hat{H}}$, i.e.,
\begin{eqnarray}
	G_{\alpha\gamma} = \langle \alpha |e^{-\tau \hat{H}}|\gamma\rangle \approx \delta_{\alpha\gamma} - \tau H_{\alpha\gamma},
\end{eqnarray}
where $\alpha$ and $\gamma$ denote different spin configurations $\{\sigma\}$. By normalizing the rows of $G$, the element of $\tilde{G}=b^{-1}G$ can be considered as the transition amplitude between configurations. Then we can employ a Markov process with the transition matrix $\tilde{G}$ to evolve the wave function $|\Phi_0\rangle$. Here $b$ is a diagonal matrix whose nonzero elements $b_{\alpha} \equiv \sum_{\gamma}G_{\alpha\gamma}$. \tred{In this study, we choose the PEPS $|\Psi_0\rangle$ as the initial trial wave function $|\Phi_0\rangle$.}

In order to improve efficiency, importance sampling is adopted to guide the sampling process. For quantum systems with sign problems, \tred{the fixed-node approximation is usually adopted in the GFMC method \cite{Haaf1995PRB, sorella2017book}}, and an appropriate guiding wave function is expected to circumvent the sign problem, in which one needs to identify the nodes of the guiding wave function and fix them during the whole evolution process \tred{of $|\Phi_0\rangle$}. For this reason, it is crucial to ensure that the guiding wave function can accurately capture the nodal information of the ground state wave function \cite{Sebastian-PRB2014, CPB2022}. As mentioned above, we choose a PEPS, $|\Psi_{0}\rangle$, as the guiding wave function in our approach. At the beginning, the configurations of the Markov chain follow a distribution characterized by \tred{$\left| \Phi_{0} \right|^2$ (in our study, it is also $\left| \Psi_{0} \right|^2$)}, and then it progressively converges to the ground state after a sufficiently long imaginary-time evolution. It is expected that the PEPS wave function can capture the nodal information of the ground state wave functions, even when $D$ is small, and then GFMC can be used efficiently to refine the wave function and calculate the observables.

\begin{figure}[H]
	\centering
	\includegraphics[scale=0.32]{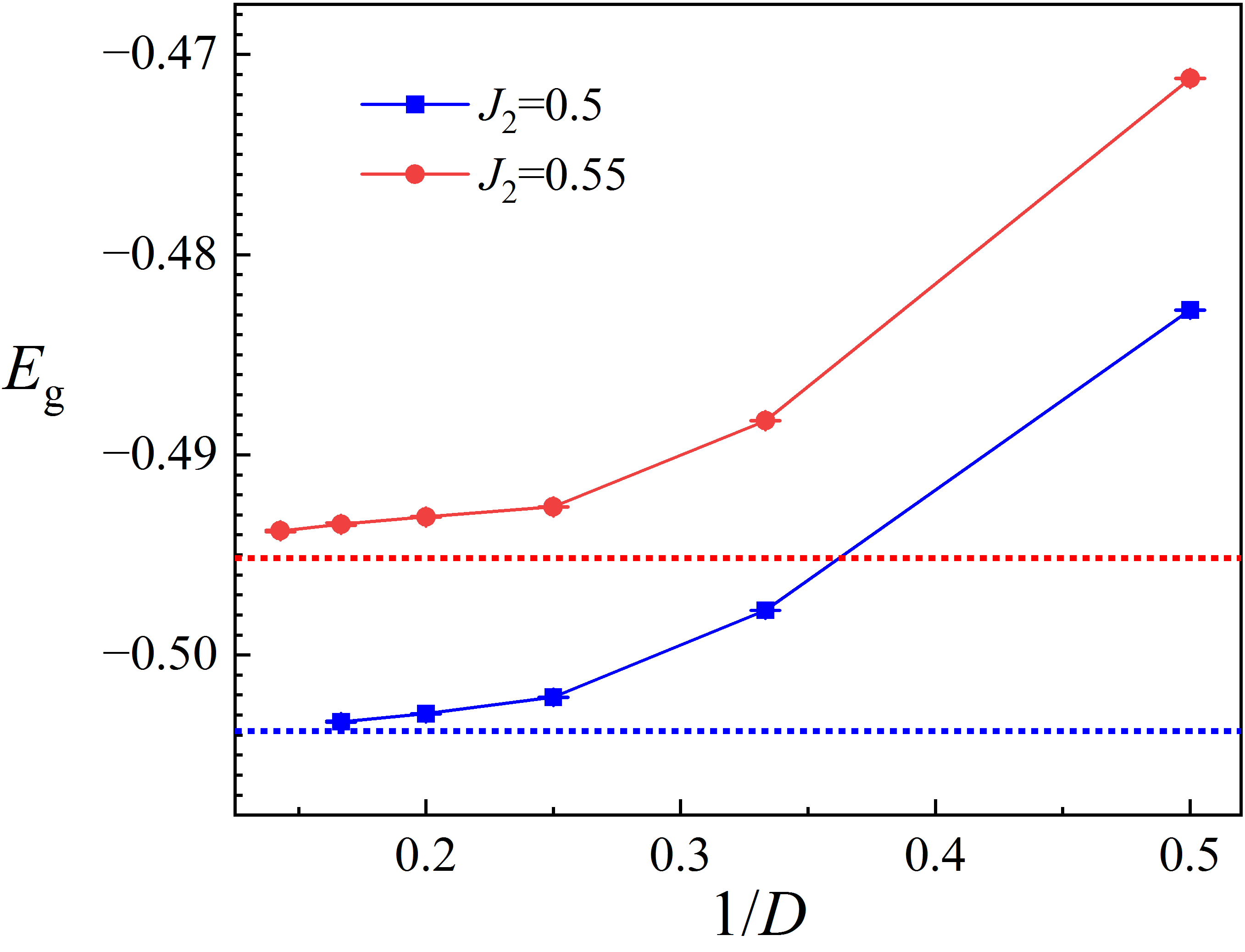}
	\caption{\tred{The ground state energy $E_g$ as a function of $D$ for a $6\times 6$ torus at $J_2 = 0.5$ (blue) and $J_2 = 0.55$ (red). The dashed lines represent the value obtained from exact diagonalization~\cite{Schulz1994Magnetic}. The extrapolated values in the large-$D$ limit can be expected to reproduce the exact value accurately \cite{Large-D}}.}
	\label{fig:Energy}
\end{figure}

{\em Ground State Energy.} To show the validity of the hybrid approach, firstly, we focus on a $L = 6$ torus, respectively with $J_2 = 0.5$ and $J_2 = 0.55$ where the frustration is known to be very strong. We report the ground state energy obtained with different bond dimensions $D$ in Tab.~\ref{tb:tb1}. It shows clearly that for both the $J_2$ values, the final energies $E_{g}$ obtained from the GFMC are indeed lower than the initial values $E_0$ provided by the PEPS wave function \tred{$|\Psi_0\rangle$}, as it should be. Though the ground state energy becomes more and more accurate when $D$ becomes larger, the GFMC can always improve the PEPS energy further. This reflects the fact that the  GFMC method goes beyond variational, and its performance relies significantly on the accuracy of the starting wave function. In this study, we keep $D \leq 7$ to balance performance and efficiency.

\begin{table*}[htbp]
    \centering
	\caption{ Energies of $J_1$-$J_2$ Heisenberg model on a $6\times6$ square lattice with $J_2 = 0.5$ and $J_2=0.55$. $E_0$ is the initial energy of the PEPS trial wave function, and $E_g$ is the energy after the GFMC optimization. The relative improvement of $E_g$ over $E_0$ is defined as $\delta =|E_{g}-E_{0}|/|E_{0}|$.}
	\renewcommand\arraystretch{1.6}
	\setlength{\tabcolsep}{5mm}
	\begin{tabular}{|c|c|c|c|c|c|c|}
		\hline
		 & \multicolumn{3}{c|}{$J_2=0.5$} & \multicolumn{3}{c|}{$J_2=0.55$} \\
		\hline
		$D$ & $E_{0}$ & $E_{g}$ & $\delta$ & $E_{0}$ & $E_{g}$ & $\delta$ \\
		\hline
		2 & -0.4817(3) & -0.48277(2) & 0.0022 & -0.4638(3) & -0.4712(1) & 0.0160 \\
		\hline
		3 & -0.4819(2) & -0.49777(3) & 0.0329 & -0.4857(2) & -0.4883(1) & 0.0054\\
		\hline
		4 & -0.4883(2) & -0.50211(7) & 0.0283 & -0.4920(1) & -0.4926(1) & 0.0012\\
		\hline
		5 & -0.4919(3) & -0.5029(1) & 0.0224 & -0.4925(4) & -0.4931(3) & 0.0012\\
		\hline
        \add{6} & \add{-0.4947(4)} & \add{-0.5033(4)} & \add{0.0172} & \add{-0.4929(5)} & \add{-0.4934(6)} & \add{0.001}\\
		\hline
        \add{7} & -- & -- & -- & \add{-0.4933(2)} & \add{-0.4939(5)} & \add{0.0012}\\
		\hline
	\end{tabular}
	\label{tb:tb1}
\end{table*}

Tab.~\ref{tb:tb3} and Tab.~\ref{tb:tb4} compare our benchmark results with existing data in the literature for $L = 6$ and $L = 10$, respectively. For $L = 6$ torus, where the exact diagonalization results are available, the hybrid approach can obtain very accurate results with even $D = 5$, which is relatively small. The hybrid approach is better in performance than variational Monte Carlo (VMC) \cite{Hu2013PRB} and convolutional neural network (CNN) \cite{Choo2019PRB}, close to DMRG \cite{Gong2014PRL} and improved restricted Boltzmann machine (RBM) method \cite{Nomura2021PRX}. \tred{In Fig.~\ref{fig:Energy}, we plot the energy $E_g$ as a function of $D$. It shows clearly that the energy can be systematically improved as $D$ becomes larger, and the extrapolated values in the large-$D$ limit can be expected to reproduce the exact value accurately \cite{Large-D}}. For $L = 10$ torus, the conclusion is similar. Even with $D = 4$, our hybrid approach can provide lower energies than the VMC \cite{Hu2013PRB} and CNN \cite{Choo2019PRB}, and the values coincide very well with those obtained by DMRG calculations \cite{Gong2014PRL}.

\begin{table}[htbp]
    \centering
	\caption{Comparison between the VMC \cite{Hu2013PRB}, DMRG \cite{Gong2014PRL}, CNN \cite{Choo2019PRB}, improved RBM \cite{Nomura2021PRX} and the PEPS-GFMC combined approach used in this paper on the $6\times6$ square lattice with periodic boundary conditions. The energies obtained from exact diagonalization (ED) are taken from Ref.~\cite{Schulz1994Magnetic}.
    }
	\renewcommand\arraystretch{1.6}
	\setlength{\tabcolsep}{5mm}
	\begin{tabular}{|c|c|c|}
		\hline
		 & $J_2=0.5$ & $J_2=0.55$ \\
		\hline
		ED & -0.503810 & -0.495178 \\
		\hline
		VMC & -0.50117(1) & -0.48992(1) \\
		\hline
		DMRG & -0.503805 & -0.495167 \\
		\hline
		CNN & -0.50185(1) & -0.49067(2) \\
		\hline
		improved RBM & -0.503765(1) & -0.495075(1) \\
		\hline
		this work (D=4) & -0.50211(7) & -0.4926(1) \\
		\hline
		this work (D=5) & -0.5029(1) & -0.4931(3) \\
		\hline
        \add{this work (D=6)} & \add{-0.5033(4)} & \add{-0.4934(6)} \\
		\hline
        \add{this work (D=7)} & -- & \add{-0.4939(5)} \\
		\hline
	\end{tabular}
	\label{tb:tb3}
\end{table}

\begin{table}[htbp]
	\caption{Comparison between the VMC \cite{Hu2013PRB}, DMRG \cite{Gong2014PRL}, CNN \cite{Choo2019PRB} and the hybrid approach used in this paper on the $10\times10$ square lattice with periodic boundary conditions.}
	\renewcommand\arraystretch{1.6}
	\setlength{\tabcolsep}{5mm}
	\begin{tabular}{|c|c|c|c|c|}
		\hline
		 & $J_2=0.5$ & $J_2=0.55$ \\
		\hline
		VMC & -0.49521(1) & -0.48335(1) \\
		\hline
		DMRG & -0.495530 & -0.485434 \\
		\hline
		CNN & -0.49516(1) & -0.48277(1) \\
		\hline
		this work (D=4) & -0.4954(6) & -0.4852(3) \\
		\hline
        \add{this work (D=5)} & \add{-0.4957(2)} & \add{-0.4853(4)} \\
		\hline
	\end{tabular}
	\label{tb:tb4}
\end{table}

\begin{figure}[H]
	\centering
	\includegraphics[scale=0.32]{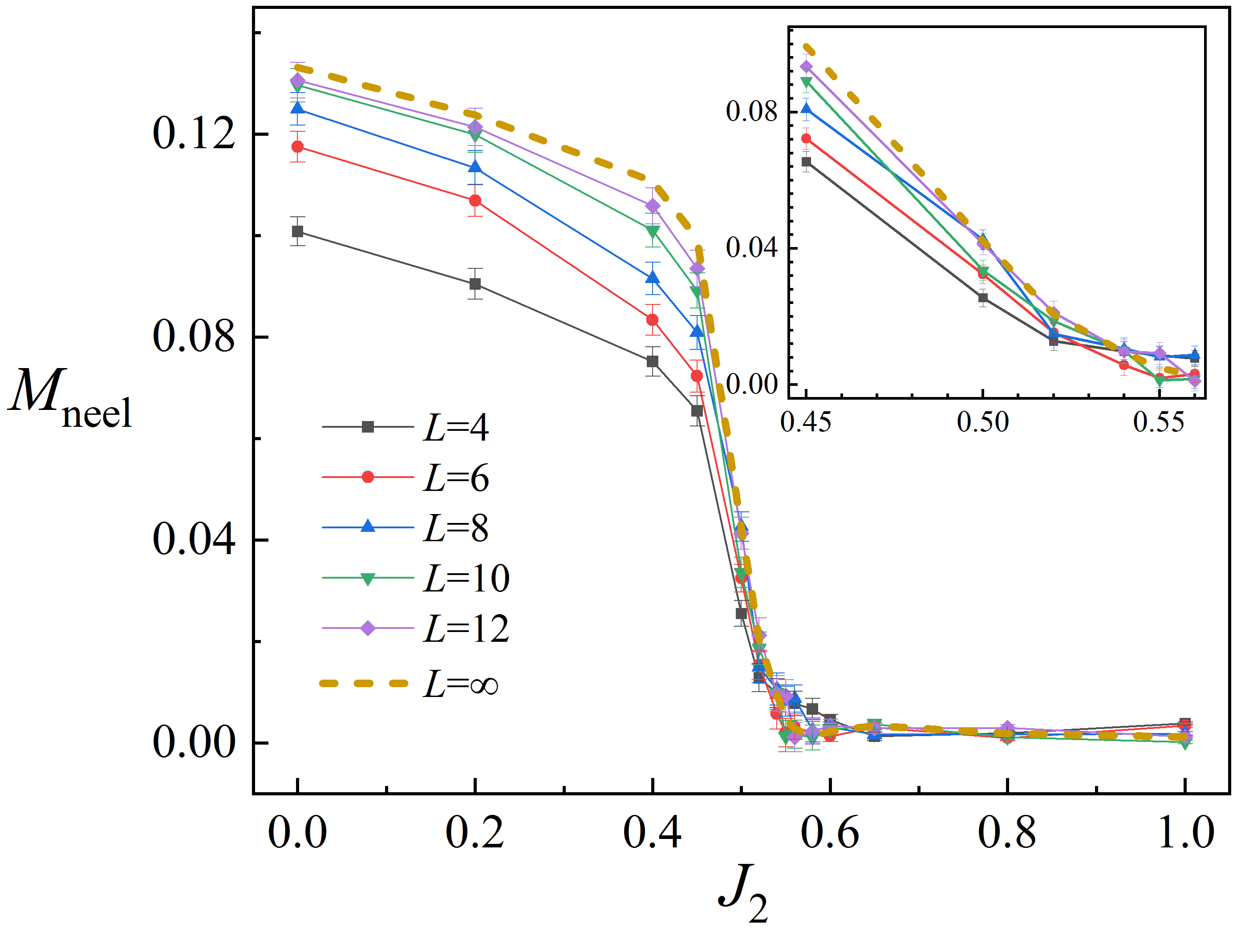}
	\caption{\tred{The relevant local order parameter $M_{\textrm{neel}}$ to the antiferromagnetic Neel phase, for $D = 4$ and a series of lattice sizes. The orange dashed line shows a direct extrapolation to the thermodynamic limit.}}
	\label{fig:Mneel}
\end{figure}

\begin{figure}[H]
	\centering
	\includegraphics[scale=0.32]{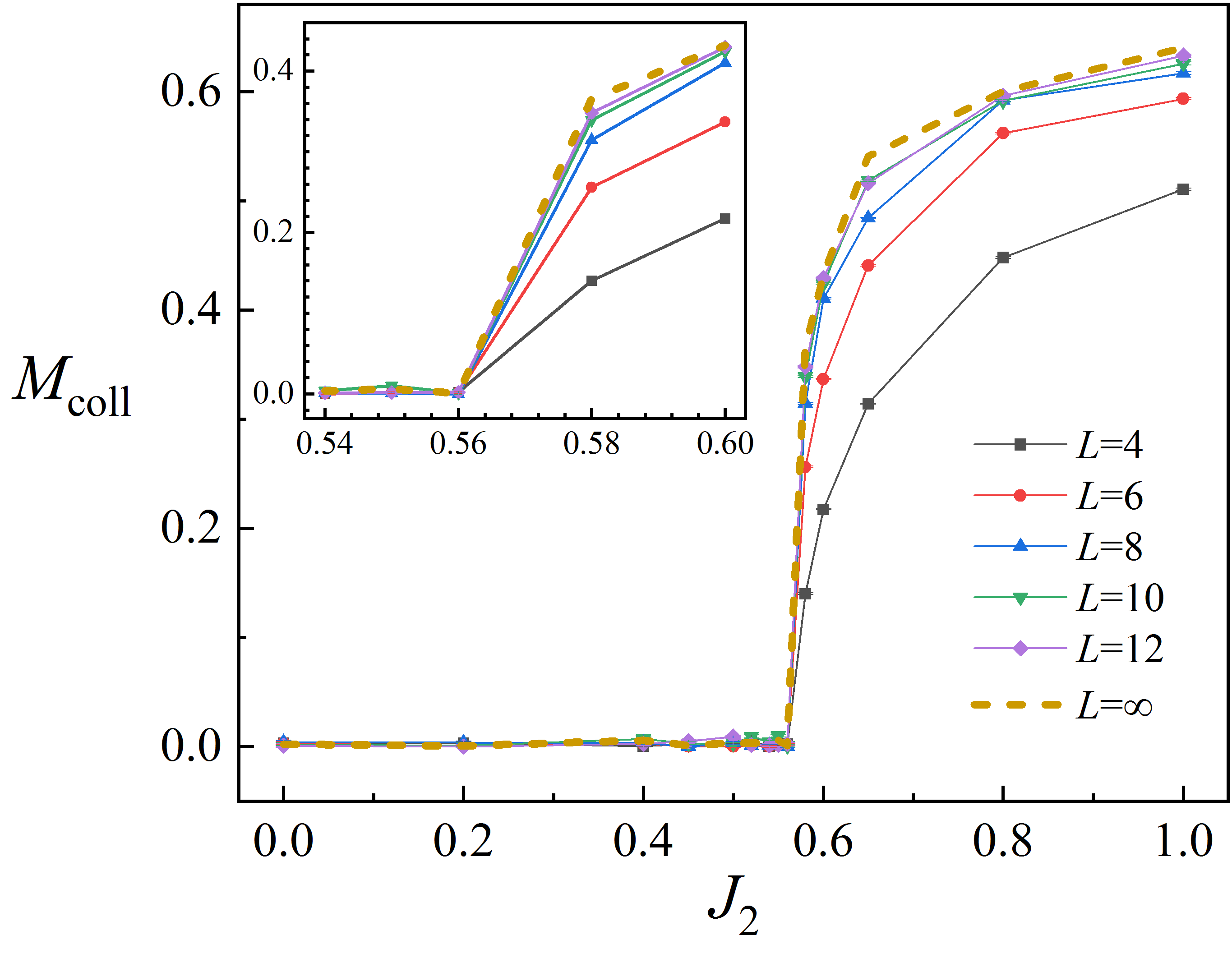}
	\caption{\tred{The relevant local order parameter $M_{\textrm{coll}}$ to the antiferromagnetic collinear phase, for $D = 4$ and a series of lattice sizes. The orange dashed line shows a direct extrapolation to the thermodynamic limit.}}
	\label{fig:Mcoll}
\end{figure}

{\em Phase Diagram.} To study the ground state phase diagram, we first calculate the magnetization density that is defined in the following
\begin{eqnarray}
	M = \sqrt{\langle S^x\rangle^2 + \langle S^y\rangle^2 + \langle S^z\rangle^2},
\end{eqnarray}
where
\begin{eqnarray}
	 S^{\alpha} = \frac{1}{L^2}\sum_{ij} S^{\alpha}_{ij}(-1)^{n},   \quad\alpha=x,y,z \label{eq:Mag}
\end{eqnarray}

\begin{figure*}
	\centering
	\subfloat[$J_2=0$]{\includegraphics[width=.3\linewidth]{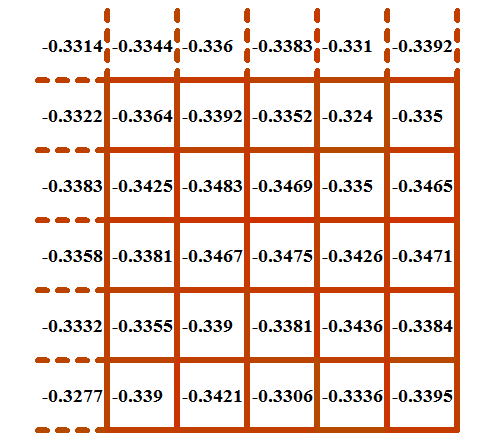}\label{fig:sub1}}
	\subfloat[$J_2=0.55$]{\includegraphics[width=.3\linewidth]{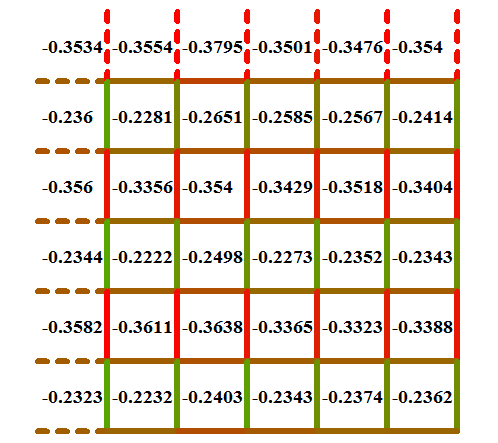}\label{fig:sub2}}
	\subfloat[$J_2=1$]{\includegraphics[width=.37\linewidth]{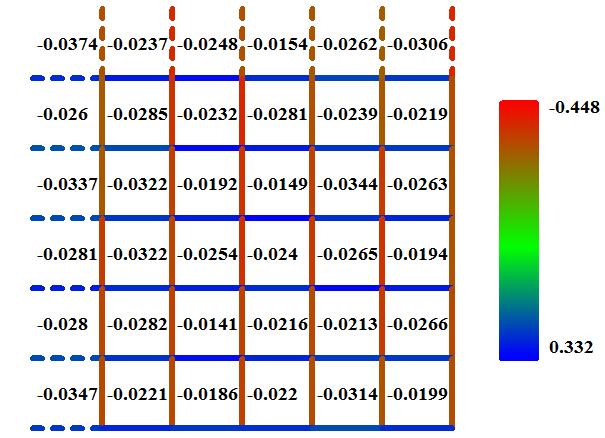}\label{fig:sub3}}
	\caption{\tred{Bond correlation for each nearest neighbor, for $D = 4$ and $L = 6$. The color indicates the strength of the bond correlation. For each square, the number in the center represents the mean of the correlations corresponding to the four sides.}}
	\label{fig:BondCorr}
\end{figure*}

Here the subscripts $(ij)$ denote that the spin is defined at the $i$-th row and the $j$-th column of the square lattice. For the \tred{antiferromagnetic Neel order parameter}, $n = i+j$ in Eq.~(\ref{eq:Mag}), while for the \tred{antiferromagnetic collinear order parameter}, $n = i$ or $n = j$, depending on how the collinear \tred{order} stretches in space. As shown in Fig.~\ref{fig:Mneel}, the Neel magnetization $M_{\textrm{neel}}$ is nonzero when $J_2$ is small and rapidly decays to roughly zero when $J_2 > 0.52$. On the other hand, Fig.~\ref{fig:Mcoll} shows that the collinear magnetization $M_{\textrm{coll}}$ is zero when $J_2$ is small, but escalates fast when $J_2 > 0.58$. In the intermediate region, about $0.52 < J_2 < 0.58$, both the two magnetizations are very small, indicating a possible nonmagnetic state.

\begin{figure}[H]
	\centering
	\includegraphics[scale=0.32]{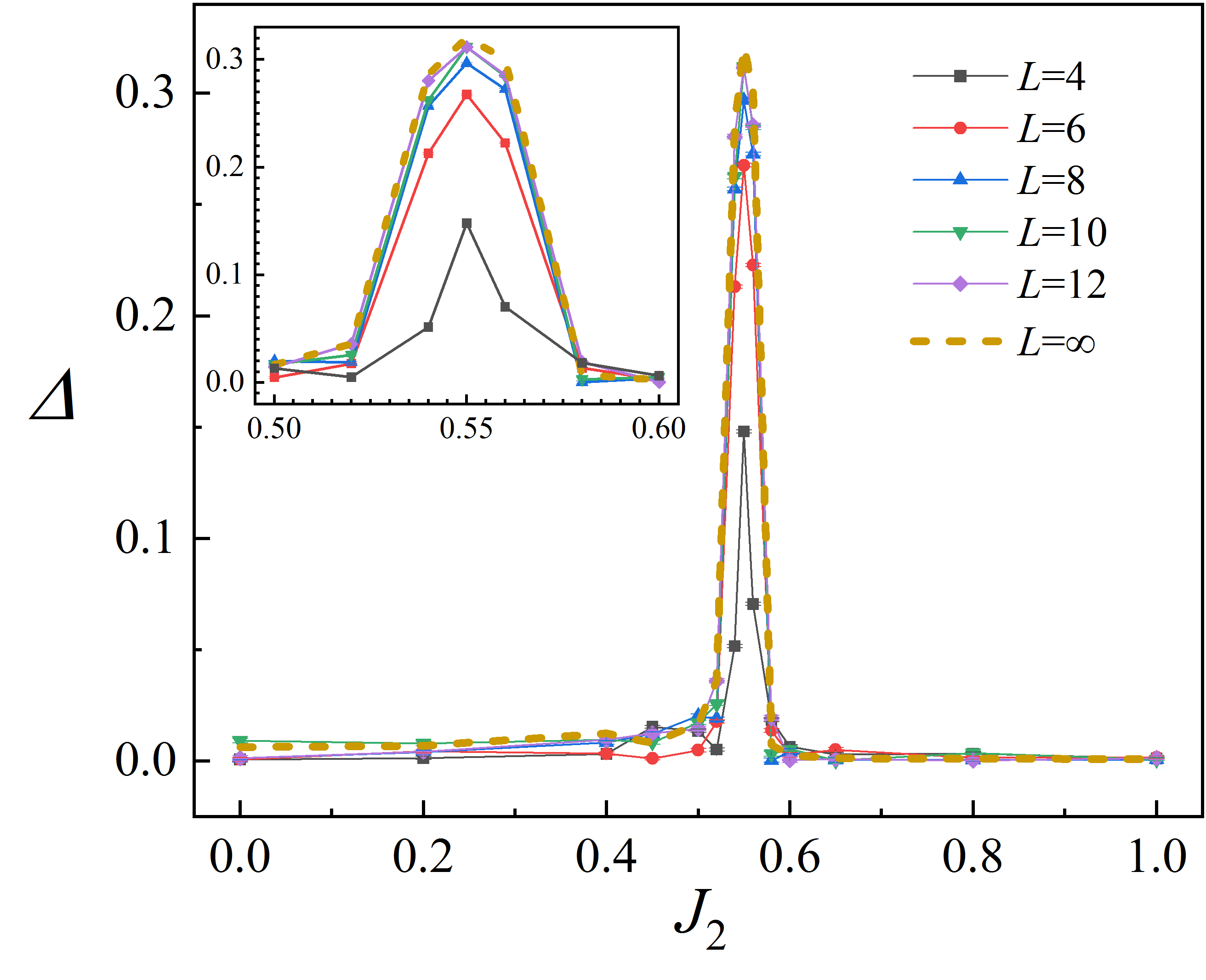}
	\caption{\tred{The relevant order parameter to the columnar VBS phase, for $D = 4$ and a series of lattice sizes. The orange dashed line shows a direct extrapolation to the thermodynamic limit.}}
	\label{fig:CVBS}
\end{figure}

To unveil the nature of the intermediate phase, we plot the bond correlation $\langle{\mathbf S_i}\cdot{\mathbf S_j}\rangle$ for each nearest neighbor in Fig.~\ref{fig:BondCorr} for $J_2 = 0.55$, where the Neel phase ($J_2 = 0$) and collinear phase ($J_2 = 1$) are also included for comparison. It shows clearly that different from $J_2=0$ (Neel phase) and $J_2=1$ (collinear phase) cases, where lattice translation symmetry for this quantity is conserved, the intermediate phase simultaneously breaks translational symmetry and rotational symmetry. In fact, this phase shows a clear columnar VBS feature along the $y$-direction, which can be characterized by a local order parameter $\Delta$ defined as
\begin{eqnarray}
	\Delta=\frac{2}{N}\left(\sum_{i\in\textrm{even},j}\langle\mathbf S_{i,j}\cdot{\mathbf S_{i+1,j}}\rangle-\sum_{i\in\textrm{odd},j}\langle\mathbf S_{i,j}\cdot{\mathbf S_{i+1,j}}\rangle\right)
\end{eqnarray}

In Fig.~\ref{fig:CVBS}, we plot its expectation value with respect to $J_2$. It shows clearly that this quantity is nonzero only in a narrow region of about $0.52 < J_2 < 0.58$, which agrees well with what we have concluded from the magnetizations.

\begin{figure}[H]
	\centering
	\includegraphics[scale=0.32]{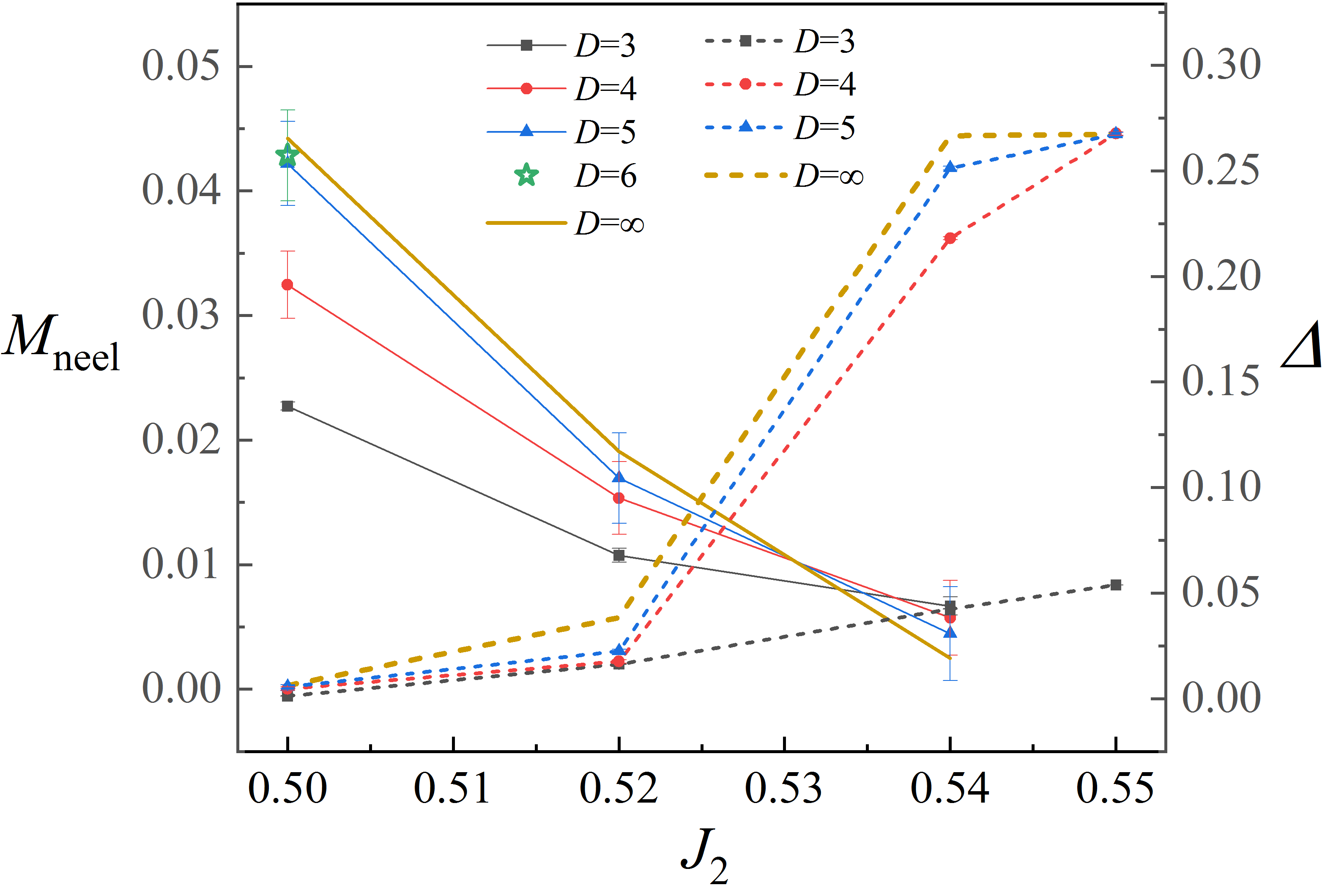}
	\caption{\tred{The Neel order parameter $M_{\text{neel}}$ (solid) and columnar VBS order parameter $\Delta$ (dashed), as functions of $D$, for a $6\times 6$ torus near the critical region $0.5\leq J_2 \leq 0.55$. The orange lines indicate direct extrapolation to the large-$D$ limit.}}
	\label{fig:OPD}
\end{figure}

\begin{figure*}[htbp]
\centering
\includegraphics[scale=0.65]{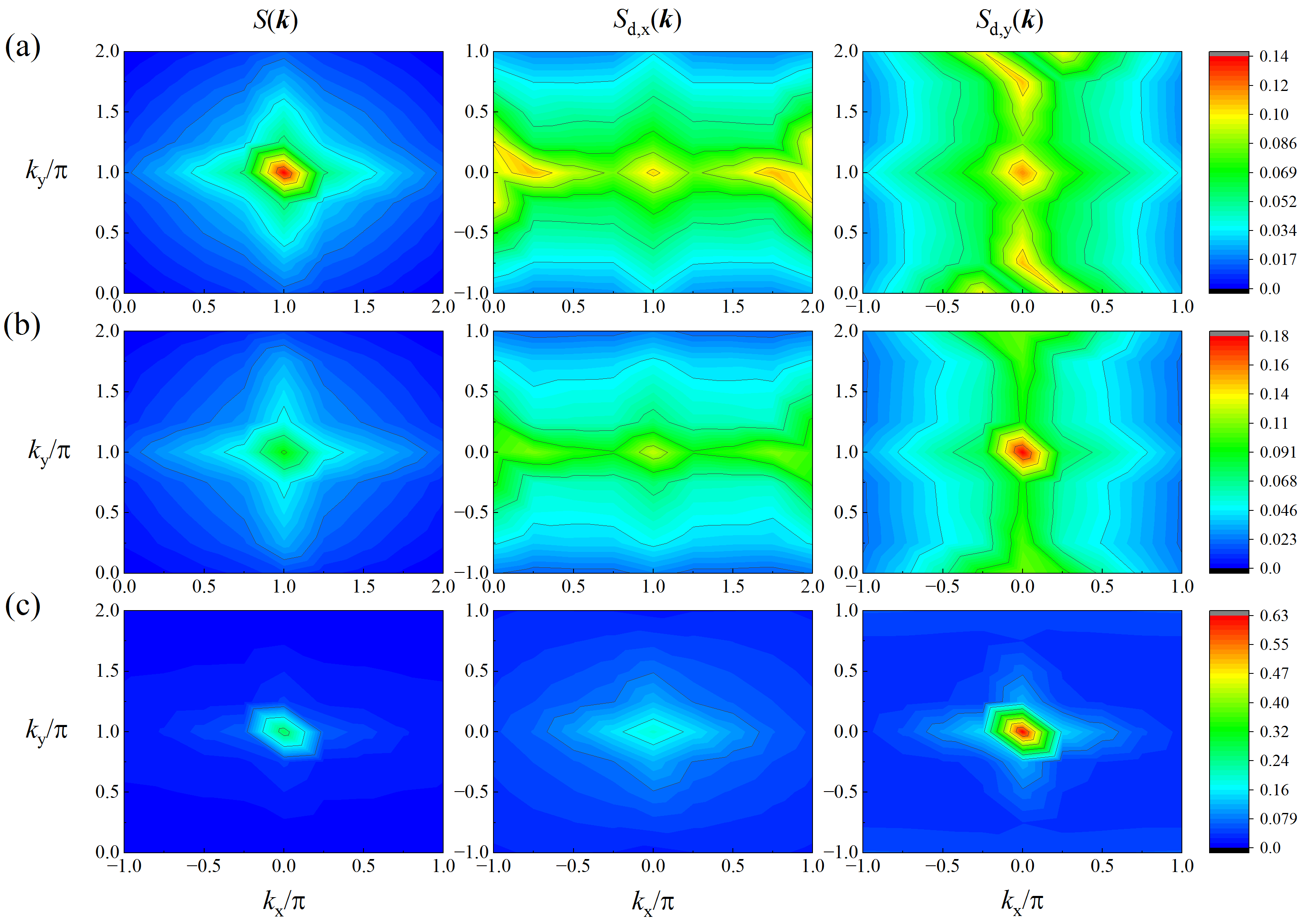}
\caption{\tred{Static structure factors near the intermediate phase, for $D = 4$ and $L = 6$. (a) $J_2 = 0.45$. (b) $J_2 = 0.55$. (c) $J_2 = 0.65$.}}
\label{fig:SF}
\end{figure*}

\tred{Apart from the results of $6\times 6$ torus, in Fig.~\ref{fig:Mneel}, Fig.~\ref{fig:Mcoll}, and Fig.~\ref{fig:CVBS}, we also plot the results for larger clusters up to $L=12$ as well as extrapolations to the thermodynamic limit. It shows that, when system size becomes larger, $M_{\text{neel}}$, $M_{\text{coll}}$ and $\Delta$ roughly become more stronger in the region of $J_2 < 0.52$, $J_2 > 0.58$, and $0.52 < J_2 < 0.58$, respectively. Results obtained for tori with different sizes give consistent conclusions.}

\tred{In Fig.~\ref{fig:OPD}, we also checked the Neel order parameter $M_{\text{neel}}$ and columnar VBS order parameter $\Delta$ for larger $D$, near the boundary between the two phases. It shows that the qualitative conclusion holds for larger $D$: when $J_2 > 0.52$, $M_{\text{neel}}$ rapidly decays to roughly zero and $\Delta$ rises fastly, that is, the columnar VBS order gradually establishes in this region. This verifies our expectation that the PEPS $|\Psi_0\rangle$ can roughly capture the nodal structure of the true ground state wave function, even when $D$ is small, and the GFMC method can do further optimization based on this nice feature. Moreover, it shows that when $D$ becomes larger, the Neel and columnar VBS phases become more stable in the $J_2 < 0.52$ and $0.52 < J_2 < 0.58$, respectively. Therefore, together with the finite-size analysis, we find a columnar VBS phase in the intermediate region about $0.52 < J_2 < 0.58$, and do not observe the possible quantum spin liquid phase between the Neel phase and VBS phase.}

To further justify our result, we also calculated the static structure factor $S(k)$ for spin-spin correlation
\begin{equation}
	\label{eq:spin}
	S\left( {\mathbf k} \right) = \frac{1}{N^2}\sum\limits_{i,j}{\left\langle{\mathbf S_i}\cdot{\mathbf S_j}\right\rangle}{e^{i\mathbf k \cdot \left( {{{\mathbf r}_i} - {{\mathbf r}_j}} \right)}},
\end{equation}
and dimer-dimer correlation
\begin{equation}
	\label{eq:dimer}
	S_{d,\alpha}\left( {\mathbf k} \right) = \frac{1}{N^2}\sum\limits_{i,j}\big[\langle {D}_{i,\alpha}{D}_{j,\alpha}\rangle - \langle {D}_{i,\alpha}\rangle\langle {D}_{j,\alpha}\rangle\big]
	{e^{i\mathbf k \cdot \left( {{{\mathbf r}_i} - {{\mathbf r}_j}} \right)}},
\end{equation}
where ${D}_{i,\alpha} = {\mathbf S_{i}}\cdot{\mathbf S_{i+\hat{\alpha}}}$ is the bond correlation defined for the $i$-th site, where $\alpha = x, y$ denotes the two orientations for square lattice.

The structure factors for $J_2 = 0.45, 0.55, 0.65$ are shown respectively in Fig.~\ref{fig:SF}(a), Fig.~\ref{fig:SF}(b) and Fig.~\ref{fig:SF}(c) in detail. Fig.~\ref{fig:SF} shows that when $J_2 = 0.45$, $S({\mathbf k})$ displays a clear peak at $(\pi, \pi)$. Though $S_{d,x}({\mathbf k})$ also develops weak peaks along $k_y = 0$, it does not show clear $k_x$ dependence. It is likewise with $S_{d,y}({\mathbf k})$. This means that when the ground state is in an antiferromagnetic Neel order, the columnar VBS order is not well-established. When $J_2 = 0.55$, the $S({\mathbf k})$ still peaks at $(\pi, \pi)$, but its magnitude is much smaller. Meanwhile, $S_{d,y}({\mathbf k})$ shows a clear peak at $(0,\pi)$. This means that the Neel order is significantly suppressed and meanwhile the columnar VBS order in $y$-direction is well established and dominates in this phase. When $J_2 = 0.65$, as shown in Fig.~\ref{fig:SF}(c), the $S({\mathbf k})$ peaks at $(0,\pi)$, $S_{d,x}({\mathbf k})$ peaks weakly at $(0,0)$, and at the same time $S_{d,y}({\mathbf k})$ shows a clear peak at $(0,0)$. All these signatures indicate that the ground state has entered a collinear antiferromagnetic phase, and the bond correlation preserves the translation symmetry. The analysis here coincides well with the pattern sketched in Fig.~\ref{fig:BondCorr}.

\begin{figure}[H]
\centering
\includegraphics[scale=0.32]{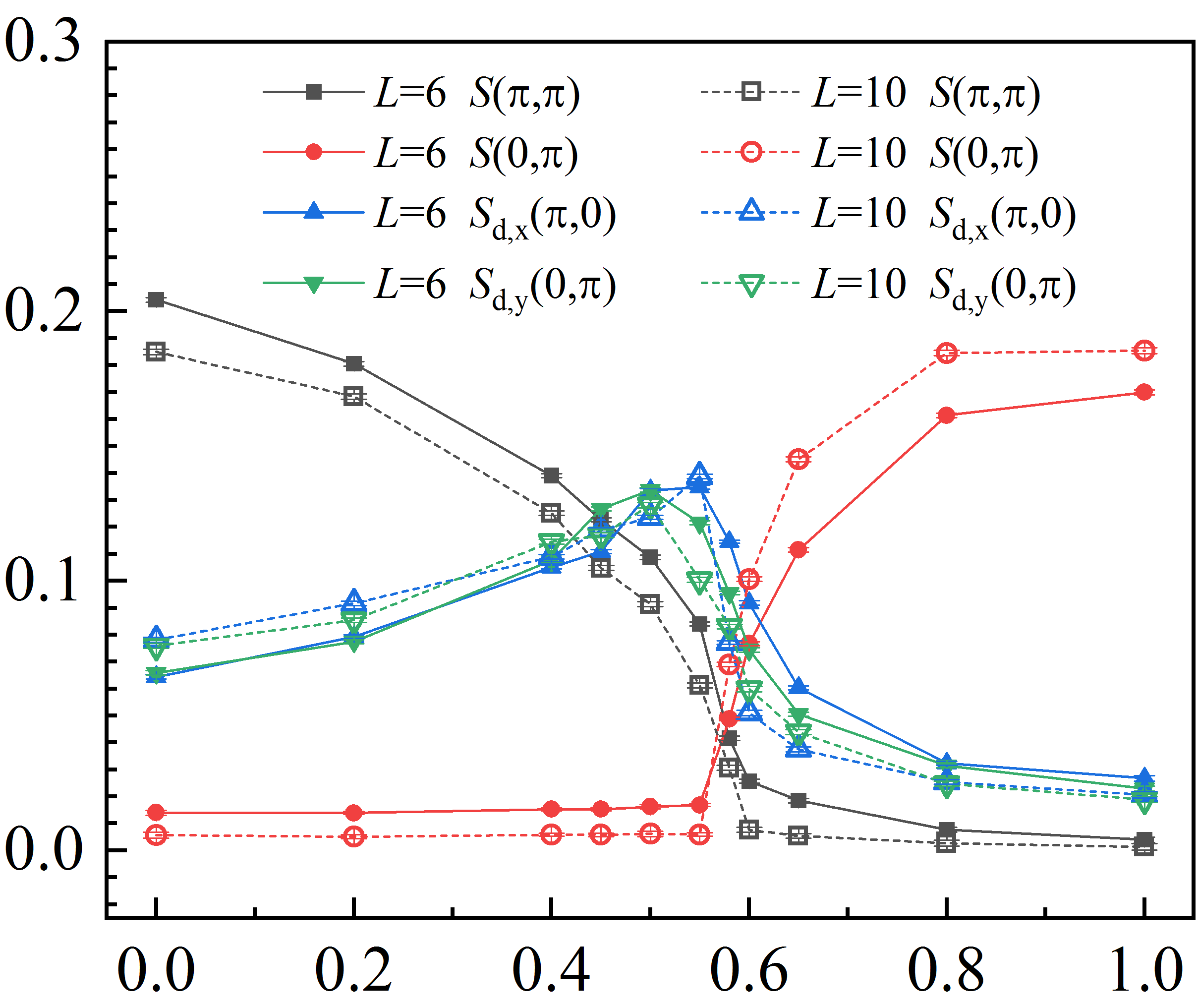}
\caption{\tred{The structure factors as functions of $J_2$, for $D = 4$, on $6\times 6$ (solid) and $10\times 10$ (dashed) tori.}}
\label{fig:L6L10}
\end{figure}

The above evolution picture can also be identified in Fig.~\ref{fig:L6L10}, where we plot the value of the characteristic peaks, i.e., $S(\pi,\pi)$ for Neel order, $S(0,\pi)$ for collinear order, $S_{d,x}(\pi,0)$ and $S_{d,y}(0,\pi)$ for columnar VBS order, concerning $J_2$. \tred{To show finite-size effect, we plot the results of $L = 6$ and $L = 10$ together for comparison.} It shows clearly that as $J_2$ becomes larger, the Neel order becomes weaker and weaker while the collinear order gradually develops. The columnar VBS order is established in the intermediate phase, and is consistent with the order parameter analysis. \tred{The results from the two tori differ only in small values, and the qualitative conclusion remains the same. More details about the finite-size analysis on the structure factors can be found in App.~III.}

{\em Summary and discussion.} In this study, we propose a hybrid approach that combines the GFMC method with the PEPS ansatz to investigate the ground state phase diagram of the frustrated $J_1$-$J_2$ model on a square lattice. By utilizing the preliminary PEPS state obtained through automatic differentiation \cite{dTRG-PRB2020, WL2019PRX} as a guiding wave function for GFMC, the hybrid approach significantly enhances the accuracy of the ground state energy. This supports the argument that the PEPS can accurately characterize the nodal structure of quantum frustrated systems, as also evidenced in previous studies \cite{Sebastian-PRB2014, MPQin-PRB2020}. Furthermore, the GFMC can  efficiently give the physical observables, including the correlation functions \cite{CPB2022}. Our results, obtained from \tred{system sizes up to $L = 12$ and bond dimensions up to $D = 7$}, including various order parameters and static structure factors, reveal the presence of a possible columnar VBS phase in the intermediate regime around $0.52 \textless J_2 \textless 0.58$. The conclusion is consistent with the previous studies by Schwinger boson approach \cite{PhysRevB.44.12050}, series expansion \cite{singh1999dimer}, and the more recent symmetric infinite PEPS \cite{haghshenas2018u}. \tred{Though we cannot fully exclude the possibility of the existence of a narrow quantum spin liquid phase \cite{wang2018critical,Nomura2021PRX}, we tend to believe that there is no such state between the Neel phase and the VBS phase from the finite-size and finite-D analysis on the order parameters.}

The hybrid approach capitalizes on the PEPS' capability of characterizing the nodal structure of quantum many-body states and the full parallelizability of Monte Carlo sampling. \tred{Specifically, given the PEPS guiding wave function with bond dimension $D$, the hybrid method can further optimize it with computational cost scales as $nD^6$, where $n$ is the number of spin configurations sampled in the GFMC, and $D^6$ comes from the evaluation of $\langle\{\sigma\}|\Psi\rangle$ for a given $\{\sigma\}$. While PEPS with a larger $D$ is expected to capture a more accurate nodal structure, $n$ should increase with $L$ but can be parallelized completely.} Consequently, more computational resources directly translate to enhanced results. For instance, employing larger $D$, e.g., through the nested tensor network technique \cite{NTN2017, iTEBD2023}, and more accurate trial states for frustrated systems with finite size, e.g., using other tensor network ansatz like projected entangled simplex state \cite{PESS2014} and energy minimization at finite size directly, increasing system size, and augmenting the number of samples can further refine the outcomes and provide a clearer understanding of the intermediate phase, such as the exact boundaries of the VBS phase and the nature of the transition between this VBS and the Neel phase \cite{DQCP}. We would like to leave these topics as future pursuits.

{\em \tred{Acknowledgments.}} Computational resources used in this study were provided by the National Supercomputer Center in Guangzhou with Tianhe-2 Supercomputer and the Physical Laboratory of High-Performance Computing in Renmin University of China. The first two authors (He-Yu Lin and Yibin Guo) contribute equally to this work. This work was supported by the National R\&D Program of China (Grants No. 2023YFA1406500), the National Natural Science Foundation of China (Grants No. 11934020 and No. 12274458) and the Innovation Program for Quantum Science and Technology (2021ZD0302402).

\section*{Appendix}
\tred{In this appendix, we provide more background on the tensor network states, more details on the guiding wave function preparation, and more finite-size analysis of the static structure factors.}
\section*{I. More background on tensor network states}
The PEPS wave function, a cornerstone of our research, is a widely used tensor network state in the study of quantum many-body physics \cite{verstraete2008matrix,Mont2018,orus2019tensor,cirac2021matrix,banuls2023tensor,xiang2023density, orus2014practical, Roman2014EPJB}. The two important features of PEPS and many other tensor network states are area-law scaling of the entanglement entropy and the absence of negative sign problem.

The area-law scaling originates from its specific dense structure \cite{orus2014practical}, i.e., any bipartition of the physical degrees of freedom in a PEPS will inevitably cut multiple links/bonds whose number is proportional to the system size, and this is quite different from other one-dimensional wave function ansatz like matrix product state.

The general statement that the tensor network state is free of the sign problem \cite{verstraete2008matrix,Roman2014EPJB,banuls2023tensor,xiang2023density} stems from the fact that tensor network algorithms are usually based on the idea of the renormalization group, and the concept of probability of a given configuration, like in quantum Monte Carlo, is not touched directly. For example, when tensor networks are used to study a quantum lattice model, the main task is to determine the tensor network state representation of the target quantum state \cite{verstraete2004renormalization,orus2014practical}, such as the ground state, and the strategies to achieve this include energy minimization or imaginary-time evolution. While performing the strategies and later in the expectation value calculations, the work needs to be done is to contract tensor networks approximately through some renormalization-group-based techniques, such as boundary matrix product state or corner transfer matrix renormalization group \cite{orus2014practical,xiang2023density,LXF2022CPL}, etc. In the entire process mentioned above, the concept of probability is not touched at all. Even in the imaginary-time evolution, which can be regarded as the counterpart of the path integral, the only thing one needs to do is to update the wave function by variational approaches or some singular-value-decomposition-based techniques \cite{JHC2008PRL} instead of evaluating $e^{-\tau H}$ itself. Therefore, in this sense, the problem of negative probability for a given basis is avoided completely, and it is similar to the case of variational Monte Carlo or any other wave-function-based methods. That is why it is generally said that the tensor network state is free of sign problems. There are many successful applications of TNS in systems that have sign problems for Monte Carlo, such as kagome spin liquid \cite{PESS2014}, Shastry-Sutherland model \cite{SSM20132023PRB}, lattice gauge theory with finite density \cite{QED2021TNS}, Hubbard \cite{Hubbard} and $t$-$J$ model \cite{tJmodel}, etc.

The bond dimension $D$ is an important hyper-parameter to control the number of variational parameters in the tensor network state ansatz. For example, for the PEPS defined on a $L\times L$ torus like in Fig.~\ref{fig:PEPS}, if no translational symmetry is used, then the total number of parameters is $L^2D^4d$, where $d$ is the dimension of the local Hilbert space extended by $\sigma_i$.

\begin{figure}[H]
	\centering
	\includegraphics[scale=1.2]{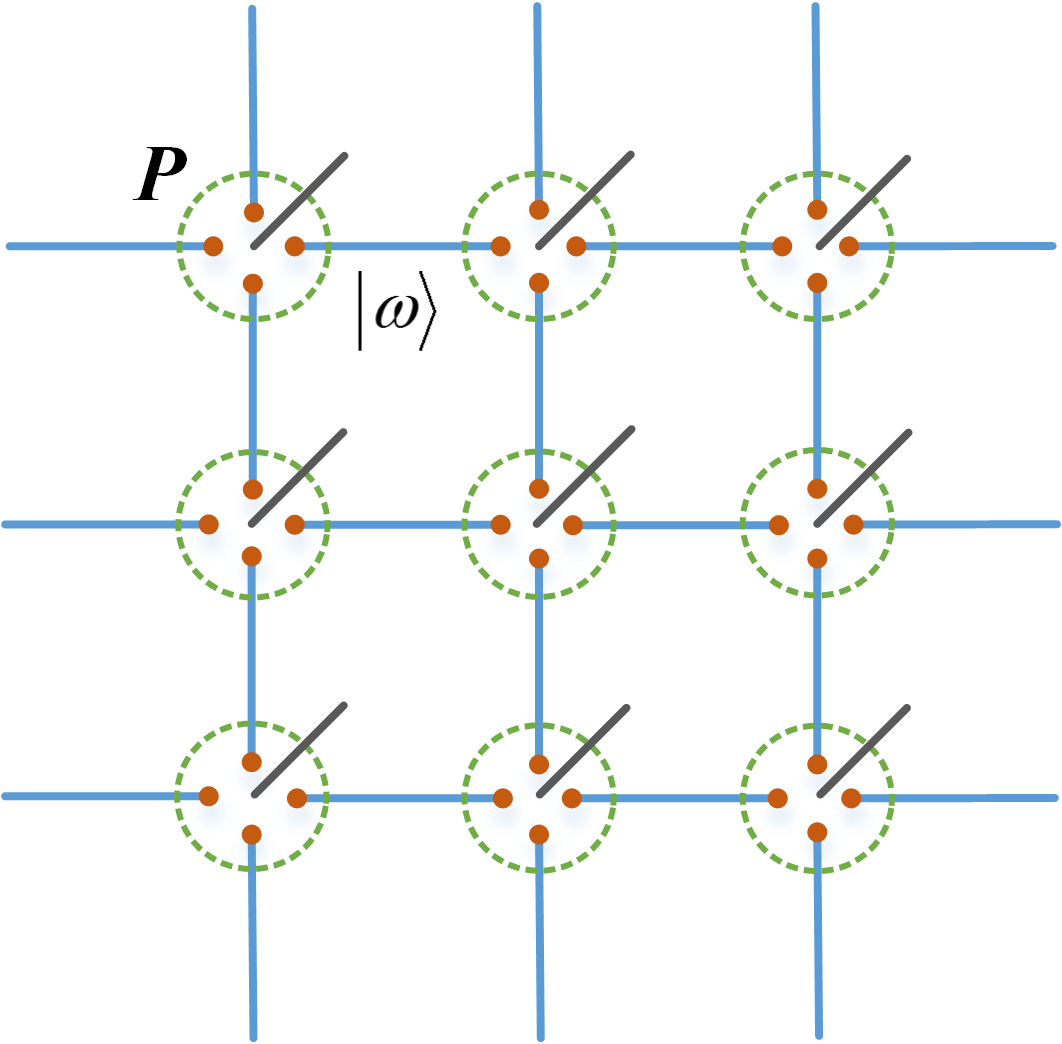}
	\caption{An interpretation of the PEPS wave function constructed on a square lattice. The red dots denote the auxiliary virtual particles, the blue lines denote maximally entangled paired state $|\omega\rangle$, and the black lines denote local physical degrees of freedom. The green dashed circles denote projectors $\textit{\textbf{P}}$ defined at each site, which maps the space expanded by the four virtual particles to the physical space. More details can be found, e.g., in Ref.~\cite{verstraete2004renormalization,Vidal2004PRL}.}
	\label{fig:PEPSdef}
\end{figure}

Physically, the PEPS wave function can be understood in terms of maximally entangled states of some auxiliary systems, as originally proposed in Refs.~\cite{verstraete2004renormalization, Vidal2004PRL}. The idea is illustrated in Fig.~\ref{fig:PEPSdef}. Firstly, one can arrange four auxiliary virtual particles around each lattice site, and let every two virtual particles on the same link form a maximally entangled paired state, e.g., $|\omega\rangle = \sum_{i=1}^D|i,i\rangle$, where $|i=1,2,...D\rangle$ characterizes the specific quantum state of the virtual particles. Then the final state is obtained by applying a projector (namely linear map) $\textit{\textbf{P}}$ at each lattice site to map the space extended by the four virtual particles to the physical Hilbert space. Therefore, in this sense, $D$ actually represents the dimension of the auxiliary virtual systems.
Generally, the PEPS ansatz can be more accurate when $D$ becomes larger, but unfortunately, the computational cost of determining and evaluating the state can scale as $D^{12}$ and increase extremely fast \cite{orus2014practical, NTN2017}. Therefore, as mentioned in the main text, one needs to balance the performance and the cost.

\section*{II. Preparation of $|\Psi_0\rangle$}
This section provides more details about the preparation of guiding PEPS wave function $|\Psi_0\rangle$. As mentioned in the main text, the PEPS ansatz is optimized for systems in the thermodynamic limit for simplicity. To be specific, we mainly follow the procedures below:

\begin{figure}[H]
	\centering
	\includegraphics[scale=1]{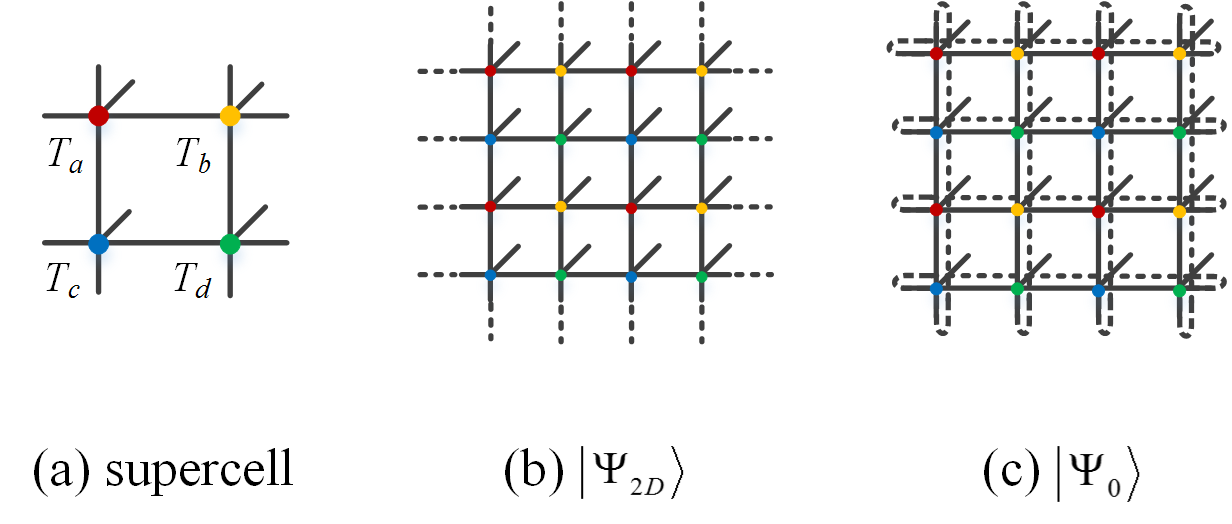}
	\caption{A sketch of the supercell (a) used to construct the PEPS wave function $|\Psi_{2D}\rangle$ in the thermodynamic limit (b) and the wave function $|\Psi_{0}\rangle$ on a $L\times L$ torus (c). Here, we use $L = 4$ as an illustration for simplicity. }
	\label{fig:WfPrepare}
\end{figure}

(1) Choose a supercell with size $2\times 2$, the smallest size necessary to distinguish the four possible phases of this model. As illustrated in Fig.~\ref{fig:WfPrepare}(a), this means there are four distinct tensors in the supercell (denoted as different colors), say $\{T_a, T_b, T_c, T_d\}$, each of which is a tensor with shape $D\times D\times D\times D\times 2$ and initialized arbitrarily.

(2) The supercell is duplicated and arranged periodically to construct the PEPS ansatz $|\Psi_{2D}\rangle$ in the thermodynamic limit. Here, the subscript '2D' addresses the thermodynamic limit of a two-dimensional lattice. In this case, only four different local tensors are constructing $|\Psi_{2D}\rangle$, i.e., $\{T_a, T_b, T_c, T_d\}$. See Fig.~\ref{fig:WfPrepare}(b).

(3) Find the optimal $\{T_a, T_b, T_c, T_d\}$ which can minimize the ground state energy $E = \frac{\langle\Psi_{2D}|H|\Psi_{2D}\rangle}{\langle\Psi_{2D}|\Psi_{2D}\rangle}$. This can be achieved by optimizing them from their initial values through the gradient-based optimization method, such as the L-BFGS quasi-Newton method \cite{LBFGS}, as long as their gradients, i.e., $\{\frac{\partial E}{\partial T_a},\frac{\partial E}{\partial T_b}, \frac{\partial E}{\partial T_c}, \frac{\partial E}{\partial T_d}\}$ are known. Fortunately, these gradients can be effectively obtained by the so-called automatic differentiation (AD) technique, which essentially uses the backpropagation (i.e., chain rule of derivatives) to calculate the gradients \cite{Biggs2000, Baydin2015}. Therefore, from the arbitrarily initialized $\{T_a, T_b, T_c, T_d\}$, one can calculate the energy $E$, and then the AD package \cite{ADsoft} can effectively obtain their gradients, which can be used in the L-BFGS method to update these local tensors, and $|\Psi_{2D}\rangle$ is thus updated in the direction of lower energy. This update procedure can be repeated until some convergence is reached, and then we obtain a PEPS representation $|\Psi_{2D}\rangle$ of the ground state wave function with some accuracy.

(4) When $|\Psi_{2D}\rangle$ is obtained, in order to combine with the GFMC method, we use the obtained supercell to approximately construct the ground state wave function of the same Hamiltonian but on a $L\times L$ torus. Again, this is done by duplicating the supercell and arranging them periodically on a torus \cite{CJW2011}. Then finally, the trial PEPS wave function $|\Psi_0\rangle$ is obtained and can be used as the guiding wave function of the GFMC method. See Fig.~\ref{fig:WfPrepare}(c).

The AD technique lies in the heart of the backpropagation algorithm in training neural networks \cite{Hinton1986BP}, relates closely to the second renormalization group in optimizing a tensor network \cite{dTRG-PRB2020}, thus servers as the computational engine of modern deep learning applications and differential programming tensor networks. The basic idea is the chain rule of derivative. For simplicity, we introduce notation $\vec{T}_0$ to denote the initial $\{T_a, T_b, T_c, T_d\}$ vectorized and stacked as a single vector. Suppose in order to evaluate the evaluation of the energy, a series of intermediate results $\{\vec{T}_1, \vec{T}_2, ..., \vec{T}_m\}$ are generated sequentially, e.g., in the corner transfer matrix renormalization group algorithm, then the gradient can be calculated through
\begin{eqnarray}
	\fp{E}{\vec{T}_0} = \fp{E}{\vec{T}_m}\fp{\vec{T}_m}{\vec{T}_{m-1}}\fp{\vec{T}_{m-1}}{\vec{T}_{m-2}}...\fp{\vec{T}_2}{\vec{T}_1}\fp{\vec{T}_1}{\vec{T}_0}.
\end{eqnarray}
There are well-developed AD packages \cite{ADsoft} to perform this reversed mode calculation effectively. For more details about this technique in tensor networks, one can refer to, e.g., Refs.~\cite{dTRG-PRB2020, WL2019PRX}.

\section*{III. More detailed finite-size analysis on the static structure factors}

\begin{figure}
	\centering
	\includegraphics[scale=0.32]{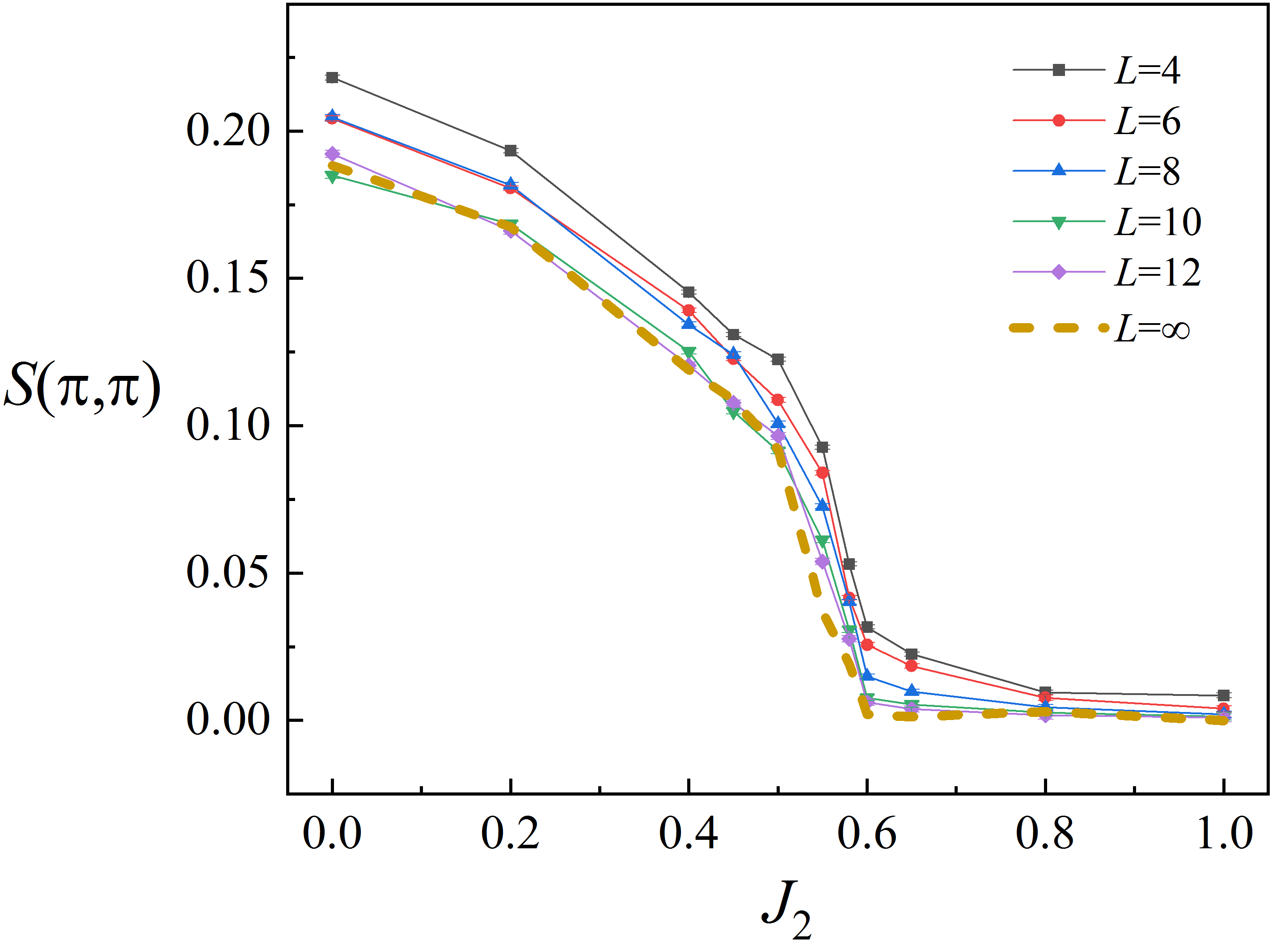}
	\caption{The relevant structure factor $S(\pi, \pi)$ to the antiferromagnetic Neel phase, for $D = 4$ and a series of lattice sizes. The orange dashed line shows a direct extrapolation to the thermodynamic limit.}
	\label{fig:Safm}
\end{figure}

\begin{figure}
	\centering
	\includegraphics[scale=0.32]{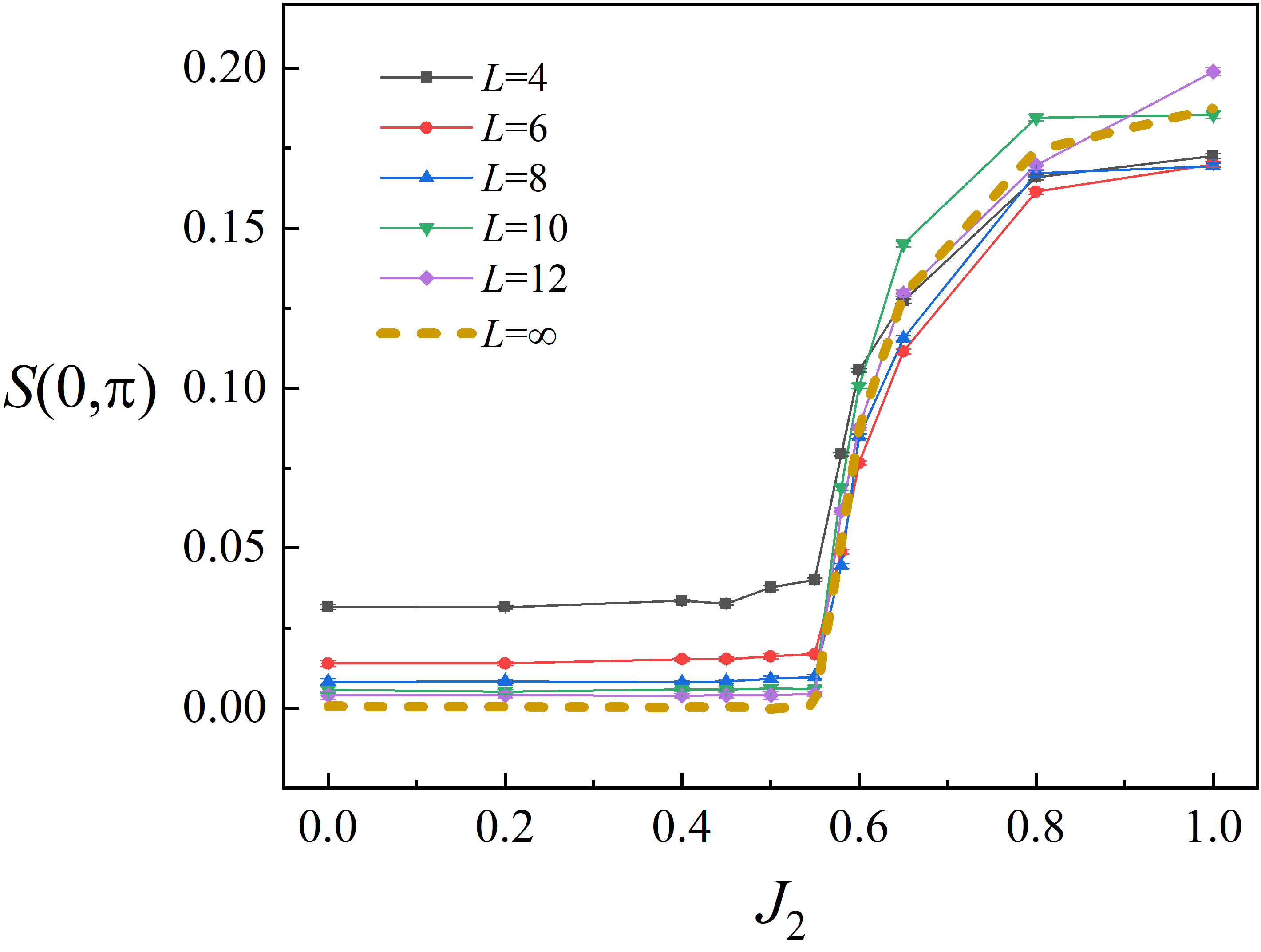}
	\caption{The relevant structure factor $S(0, \pi)$ to the antiferromagnetic collinear phase, for $D = 4$ and a series of lattice sizes. The orange dashed line shows a direct extrapolation to the thermodynamic limit.}
	\label{fig:Scol}
\end{figure}

\begin{figure}
	\centering
	\includegraphics[scale=0.32]{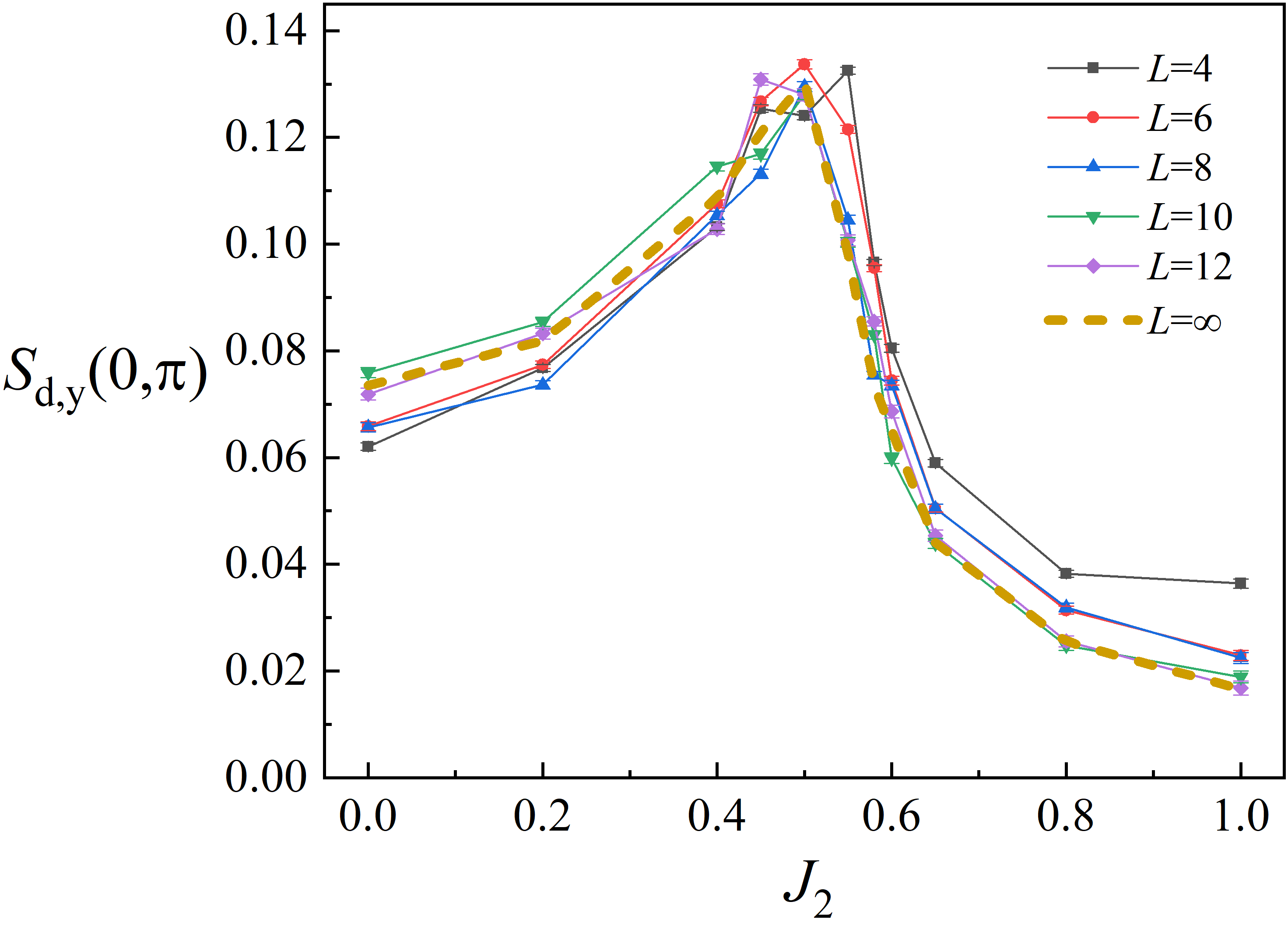}
	\caption{The relevant structure factor $S_{d,y}(0, \pi)$ to the columnar VBS phase, for $D = 4$ and a series of lattice sizes. The orange dashed line shows a direct extrapolation to the thermodynamic limit.}
	\label{fig:Svbsy}
\end{figure}

\begin{figure}
	\centering
	\includegraphics[scale=0.32]{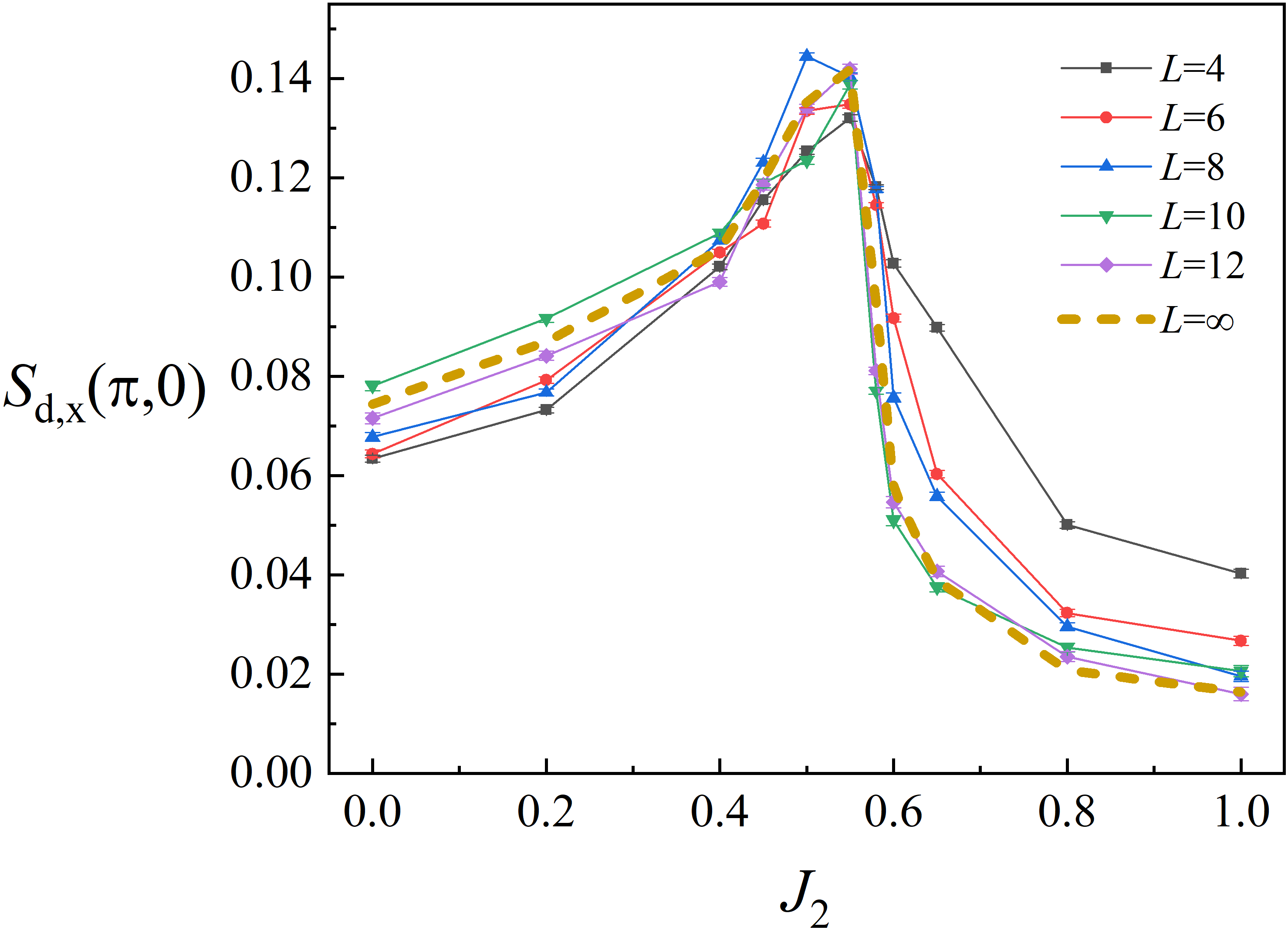}
	\caption{The relevant structure factor $S_{d,x}(0, \pi)$ to the columnar VBS phase, for $D = 4$ and a series of lattice sizes. The orange dashed line shows a direct extrapolation to the thermodynamic limit.}
	\label{fig:Svbsx}
\end{figure}

In Fig.~\ref{fig:L6L10} in the main text, we show only the data obtained on $6\times 6$ and $10\times 10$ tori in order to clarify the plot. In this section, we provide more data about the finite-size analysis for each structure factor separately. The results of $S(\pi,\pi)$ (relevant to the antiferromagnetic Neel phase), $S(0,\pi)$ (relevant to the antiferromagnetic collinear phase), $S_{d,y}(0,\pi)$ and $S_{d,x}(\pi,0)$ (relevant to the columnar VBS phase) are shown in Fig.~\ref{fig:Safm}, Fig.~\ref{fig:Scol}, Fig.~\ref{fig:Svbsy} and Fig.~\ref{fig:Svbsx}, respectively. The extrapolations to the thermodynamic limit obtained by a direct power fitting are also included. It shows that although some slight finite-size effect exists, the peak structure of all four quantities remains the same as $L$ becomes larger. Significantly, the $S_{d,y}(0,\pi)$ and $S_{d,x}(\pi,0)$ show apparent peaks in the intermediate region consistently and evidence the possible existence of a columnar VBS phase.

\bibliographystyle{apsrev4-1}
\end{document}